\documentclass[prl,twocolumn,showpacs]{revtex4}
\usepackage{graphicx}
\usepackage{amssymb,amsmath}
\usepackage{enumerate}

\newcommand{\be}{\begin{eqnarray}}
\newcommand{\ee}{\end{eqnarray}}

\begin{document}

\title{Distributed chaos and isotropic turbulence}

\author{A. Bershadskii}

\affiliation{
ICAR, P.O. Box 31155, Jerusalem 91000, Israel
}

\begin{abstract}

Power spectrum of the distributed chaos can be represented by a weighted superposition of the exponential functions which is converged to a stretched exponential $\propto \exp-(k/k_{\beta})^{\beta }$. An asymptotic theory has been developed in order to estimate the value of $\beta$ for the isotropic turbulence. This value has been found to be $\beta =3/4$. Excellent agreement has been established between this theory and the data of direct numerical simulations not only for the velocity field but also for the passive scalar, energy dissipation rate, and  magnetic fields. One can conclude that the isotropic turbulence emerges from the distributed chaos.
\end{abstract}


\maketitle

\section{Distributed chaos}

Turbulence can come from deterministic chaos \cite{swinney1}.  For dynamical systems the {\it exponential} power spectra is strong indication of chaotic dynamics \cite{fm}-\cite{sig}. The exponential spectra were also observed for the transitional and turbulent fluid motions \cite{swinney1},\cite{swinney2}. These motions were strongly anisotropic, but what can one say about isotropic turbulence? Is the isotropic turbulence related to the deterministic chaos? The exponential spectra were not observed in this turbulence even for small values of Reynolds number.   In order to approach to this problem let us start from the nature of the exponential spectra in chaotic dynamical systems. Recently certain progress was made in this direction in relation to deterministic chaos in plasmas dynamics \cite{mm}. It was suggested there that the observed exponential spectra are provided by pulses having a Lorentzian functional form \cite{mm}. Namely, for an individual Lorentzian pulse centred at time $t_0$ and having width $\tau$ the temporal shape in the complex time plain is
$$
L(t)= \frac{A}{2} \left[\frac{\tau}{\tau + i(t-t_0)} + \frac{\tau}{\tau - i(t-t_0)}\right]  \eqno{(1)}
$$ 
Then, for $N$ Lorentzian pulses the power spectrum is a sum over the residues of the
complex time poles
$$
E(\omega) \propto \left|\sum_{n=1}^{N}\exp (i\omega t_{0,n} - \omega \tau_n \right|^2  \eqno{(2)}
$$
for a narrow distribution of the pulse widths the Eq. (2) gives an exponential spectrum
$$
E(\omega) \propto \exp -(2\omega \tau)             \eqno{(3)}
$$
That is usually observed in the chaotic dynamical systems. But what if we have a broad (continuous) distribution of the pulse widths (as one can expect for isotropic turbulence)? In this case the sum Eq. (2) can be approximated by a weighted superposition of the exponential functions Eq. (3):
$$
E(\omega ) \propto \int_0^{\infty} P(\tau) \exp -(2\omega \tau) d\tau    \eqno{(4)}
$$
where $P(\tau )$ is a probability distribution of $\tau $. If the distribution is narrow and can be approximated by a delta-function then we have the exponential spectrum.

\section{Stretched exponential}
One can look at this problem as a kind of relaxation problem in a disordered medium (see for instance Ref. \cite{jon}). Namely, if one consider a global relaxation in a medium with a large number of independently relaxing species (each of which decays exponentially) and the rates of the individual exponential relaxations are different (statistically independent), then one can 
describe the global relaxation as a weighted linear superposition of the individual exponential functions in an integral form like Eq. (4). It is well known (but still not understood) that for such problems the weighted superposition of the type Eq (4) is commonly converged to a {\it stretched} exponential \cite{jon}. Let us make use of a dimensionless variable: $s=\tau/\tau_{\beta}$ (where the $\tau_{\beta}$ is a dimensional constant):
$$
E(\omega ) \propto \tau_{\beta} \int_0^{\infty} P(s) \exp -s(2\omega \tau_{\beta})~  ds    \eqno{(5)}
$$ 
then the universal stretched exponential can be written as
$$
E(\omega ) \propto \exp -(\omega/\omega_{\beta})^{\beta} \eqno{(6)}
$$
where the constant $\omega_{\beta} = (2 \tau_{\beta})^{-1}$. 
  One can solve an inverse problem and find the $P(s,\beta )$ resulting in Eq. (6). Except the case of $\beta =1/2$ the expression for $P(s,\beta)$ is rather cumbersome (see Ref. \cite{jon}). It can be useful to mention that  $P(s,\beta)$ are asymmetric Levy stable distributions. Due to the Levy-Gnedenko generalization of the central limit theorem the Levy distributions are ubiquitous in nature \cite{tsa}.
  For isotropic turbulence the frequency spectrum Eq. (6) can be transformed into wave number spectrum
$$
E(k ) \propto \exp -(k/k_{\beta})^{\beta} \eqno{(7)}
$$   
and just this spectrum should be considered as a manifestation of the distributed chaos. An example of the distributed chaos one can find in Ref. \cite{b1}.
 
\section{Asymptotic estimation of $\beta$}

\begin{figure}
\begin{center}
\includegraphics[width=8cm \vspace{-1cm}]{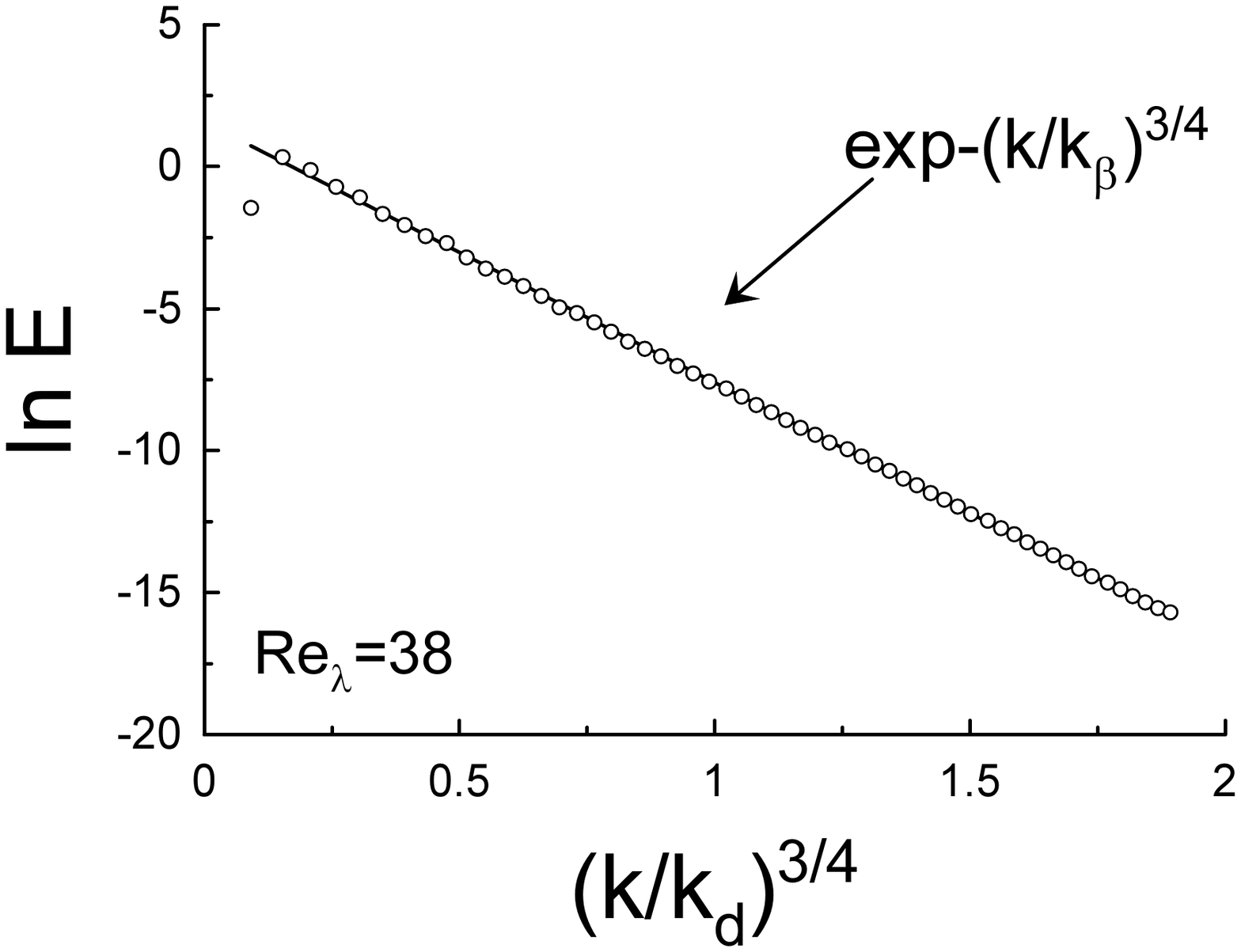}\vspace{-4cm}
\caption{\label{fig1} Three-dimensional energy spectrum from a
direct numerical simulation of isotropic steady three-dimensional turbulence \cite{gfn} for Reynolds number $Re_{\lambda} = 38$. The straight line is drawn in order to indicate a stretched exponential decay Eq. (7) with $\beta =3/4$.} 
\end{center}
\end{figure}

Asymptotic behaviour of $P(s,\beta )$ at $s \rightarrow 0$ \cite{jon}
$$
P(s,\beta ) \propto \frac{1}{s^{1+\beta/2(1-\beta)}} \exp-\frac{c}{s^{\beta/(1-\beta)}} \eqno({8)}
$$
where $c$ is a constant, can be used for an estimation of $\beta$ for the isotropic turbulence. In order to make use of this asymptotic let us consider an asymptotic of the group velocity $\upsilon (\kappa )$ of the waves (pulses) driving the distributed chaos at $\kappa \rightarrow \infty $ (do not confuse the $\upsilon(\kappa )$ with the velocity field and the wave number $\kappa$ with the wave number $k$). If at this asymptotic the group velocity has a scaling dependence on $\kappa$ 
$$
\upsilon (\kappa ) \propto \kappa^{\alpha}     \eqno{(9)}
$$
then in order to find the exponent $\alpha$ one needs to know a dimensional parameter dominating this process. It is well known that for the isotropic turbulence not only viscosity (molecular diffusion) determines the processes at $\kappa \rightarrow \infty $. Due to nonlinearity of the Navier-Stokes equations the nonlocal (long-range) interactions penetrate even into the deep dissipation range. The interplay of the nonlocality and viscosity is rather complex in this asymptotic. The opposite asymptotic $k \rightarrow 0$ for the velocity field 
is also a subject for an interplay of the long-range interactions and viscosity. Three fundamental conservation laws: energy, momentum and angular momentum, are the main source for dimensional parameters governing scaling relations in the isotropic turbulence and determining corresponding scaling exponents \cite{my}. The energy conservation law was reserved by A. Kolmogorov for the inertial range of scales where the viscosity effects are negligible \cite{my} (see also next section). Therefore, we remain with the conservation of momentum and angular momentum for the two asymptotic cases mentioned above. Both of the conservation laws were already tried in the literature for the second one from these two asymptotic. Namely, Birkhoff-Saffman and Loitsyanskii integrals (invariants) are associated with the conservation laws of momentum and angular momentum, correspondingly \cite{my}-\cite{saf}:
$$
I_n = \int_0^{\infty} r^n B_{LL} (r) dr  \eqno{(10)}
$$
where $B_{LL}$ is the longitudinal correlation function of the velocity field. For $n=2$ Eq. (10) corresponds to the 
Birkhoff-Saffman invariant and for $n=4$ to the Loitsyanskii invariant. Since conservation of the Loitsyansky integral $I_4$ depends on the type of large-scale correlations in the initial turbulent flow and, therefore, is non universal, we remain with the Birkhoff-Saffman invariant 
for determining the scaling exponent in the Eq. {(9}) from the dimensional consideration: 

$$
\upsilon (\kappa ) \simeq a~I_2^{1/2}~\kappa^{3/2} \eqno{(11)}
$$
where $a$ is a dimensionless constant.

  If $ \upsilon (\kappa ) $ has a Gaussian distribution at this asymptotic: $~ \propto \exp-(\upsilon (\kappa )/\upsilon_b)^2$, then $\kappa$ has asymptotic distribution
$$
P(\kappa ) \propto \kappa^{1/2}~\exp-\left(\frac{\kappa}{\kappa_b}\right)^3  \eqno{(12)}
$$ 
where $\kappa_b = (v_b^2/a^2I_2)^{1/3}$. 

  Taking into account that $s \propto \kappa^{-1}$ and substituting Eqs. (8) and (12) into equation
$$
P(s)~ds \propto P(k)~ dk    \eqno{(13)}
$$
we obtain $\beta = 3/4$.

\section{Comparison with the data of DNS and discussion}

 Figure 1 shows the three-dimensional energy spectrum from a
high-resolution direct numerical simulation (DNS) of homogeneous steady three-dimensional turbulence \cite{gfn} for the Taylor-scale Reynolds number $Re_{\lambda} = 38$. The scales in this figure are chosen in order to represent the Eq. (7) with $\beta =3/4$ as a straight line. The straight line is drawn in this figure in order to indicate the Eq. (7) with $\beta =3/4$. The $k_d = (\langle \varepsilon \rangle /\nu^3)^{1/4}$ (where $\nu$ is the kinematic viscosity, and $\langle \varepsilon \rangle$ is is the mean rate
of the energy dissipation \cite{my}) is the dissipation (Kolmogorov's) wavenumber. The stretched exponential covers about entire available data range and penetrates into the dissipation range $k > k_d$ without any problem. But the Kolmogorov's scale $k_d$ is still relevant to the situation because the scale $k_{\beta}$ turned out to be practically independent on the Reynolds number $Re_{\lambda}$ being represented in the terms of $k_d$: $~k_{\beta} \simeq 0.052 k_d$ (using the theory of equilibrium between chaotic and stochastic components of turbulence suggested in Ref. \cite{bb} one can obtain a theoretical estimate $~\ln(k_d/k_{\beta}) \simeq 3$). Indeed figure  2 shows the data from the same DNS but obtained at $Re_{\lambda} = 460$. The value of $k_{\beta}$ extracted from the straight line in Fig. 2 has the same value $~k_{\beta} \simeq 0.052 k_d$ (in the terms of $k_d$) as for the case $Re_{\lambda} = 38$.

\begin{figure}
\begin{center}
\includegraphics[width=8cm \vspace{-1cm}]{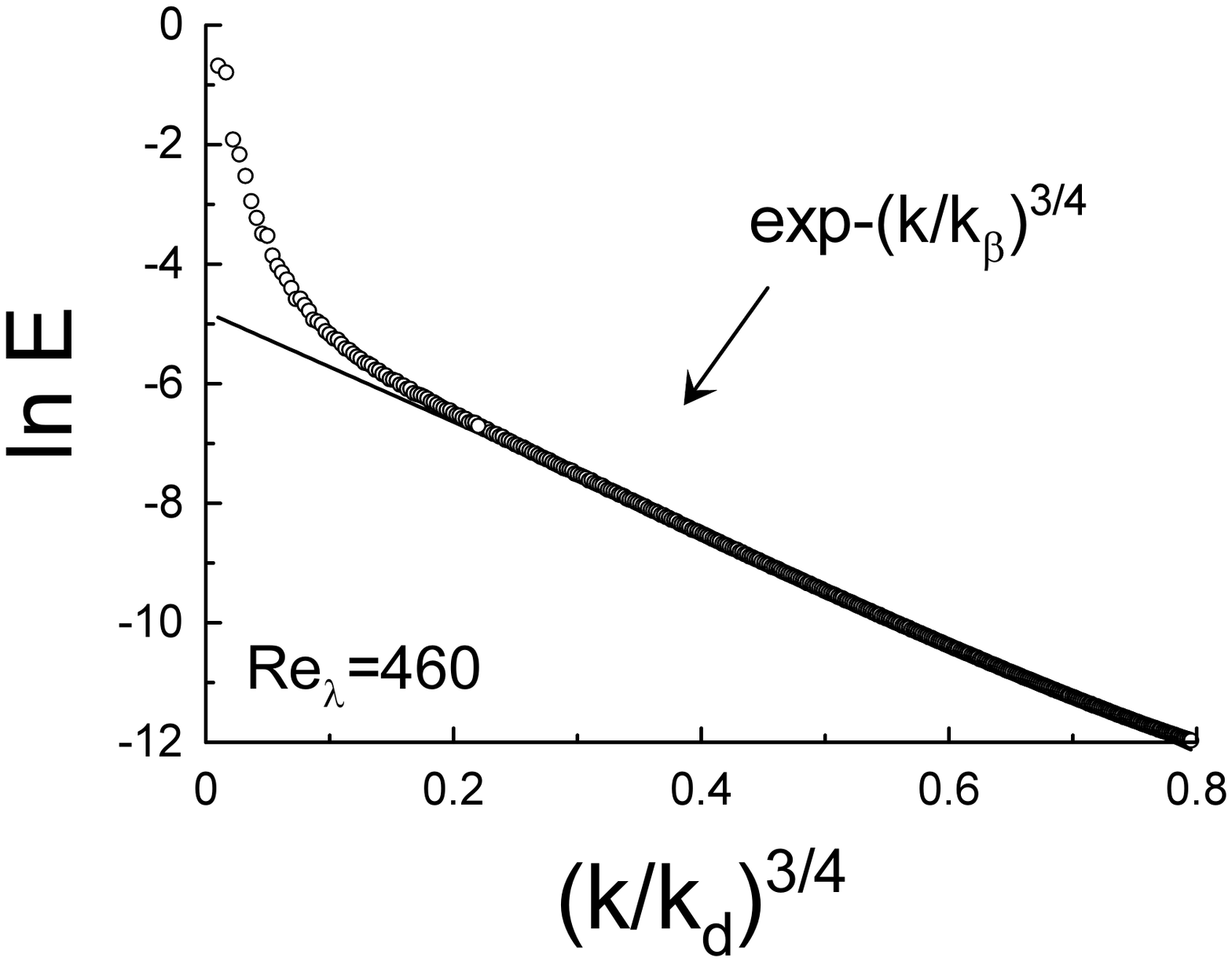}\vspace{-4cm}
\caption{\label{fig2} The same as in Fig. 1 but for Reynolds number $Re_{\lambda} = 460$}. 
\end{center}
\end{figure}

\begin{figure}
\begin{center}
\includegraphics[width=8cm \vspace{-0.8cm}]{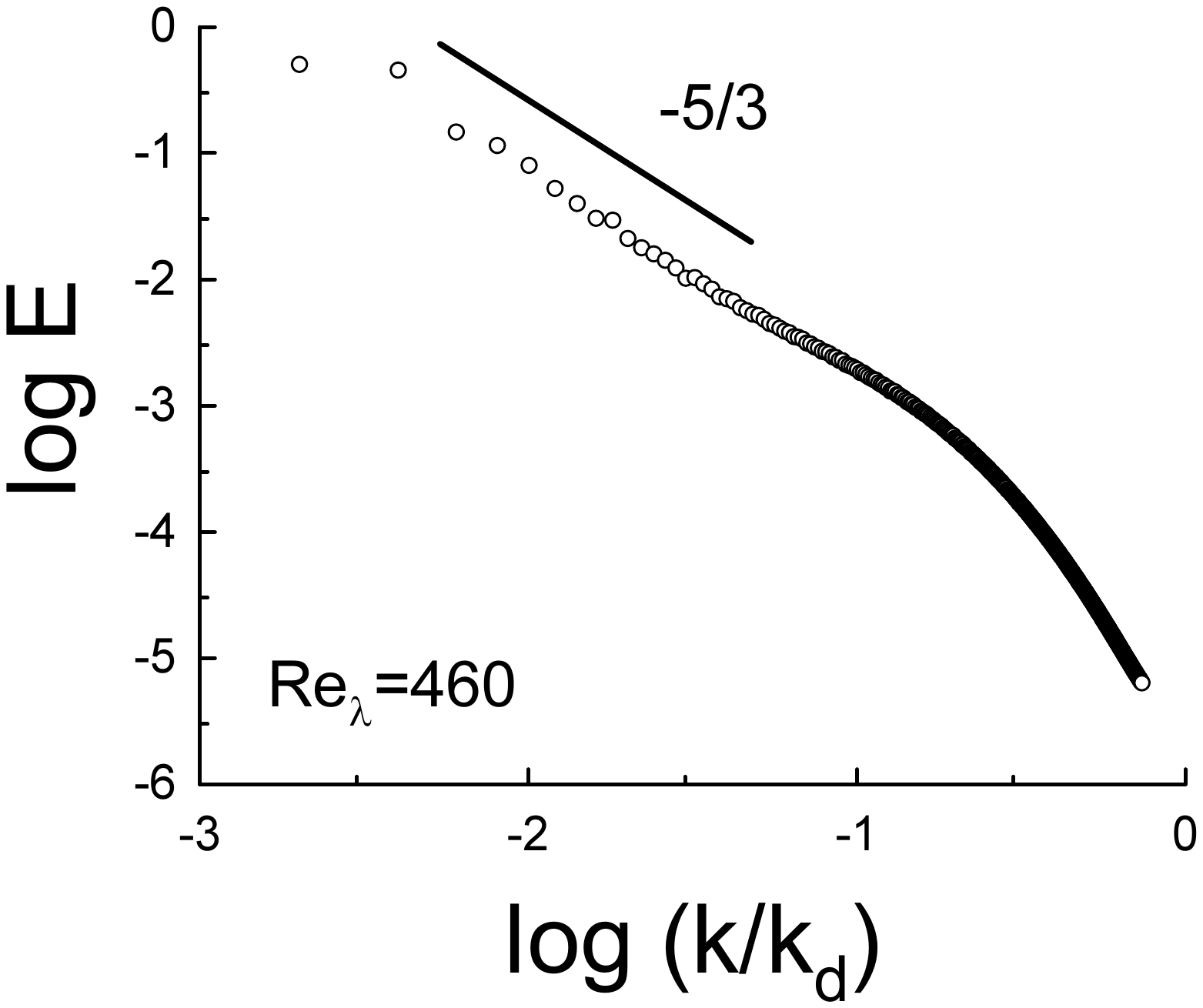}\vspace{-4.4cm}
\caption{\label{fig3} The same as in Fig. 2 but in the log-log scales}. 
\end{center}
\end{figure}

In order to show emergence of the inertial (Kolmogorov ) range adjacent to the range of the distributed chaos with the increase of $Re_{\lambda}$ we show in figure 3 the same data as in Fig. 2 but in the log-log scales (where the power-law spectrum corresponds to a straight line). Actually the range of the distributed chaos represents the range of the nonlocal interactions for this value of the Reynolds number and it is responsible here for the so-called bottleneck effect (cf., for instance, \cite{falc}-\cite{b2}). It is interesting that using the viscosity only (as a rough approximation for the governing dimensional parameter) in the asymptotic Eq. (9) one obtains an asymptotic estimate for $\beta = 2/3$. This value is only by $11\%$ smaller than the value $\beta =3/4 $ obtained above with taking into account the nonlocal interactions at the asymptotic $\kappa \rightarrow \infty$. \\

\begin{figure}
\begin{center}
\includegraphics[width=8cm \vspace{-1cm}]{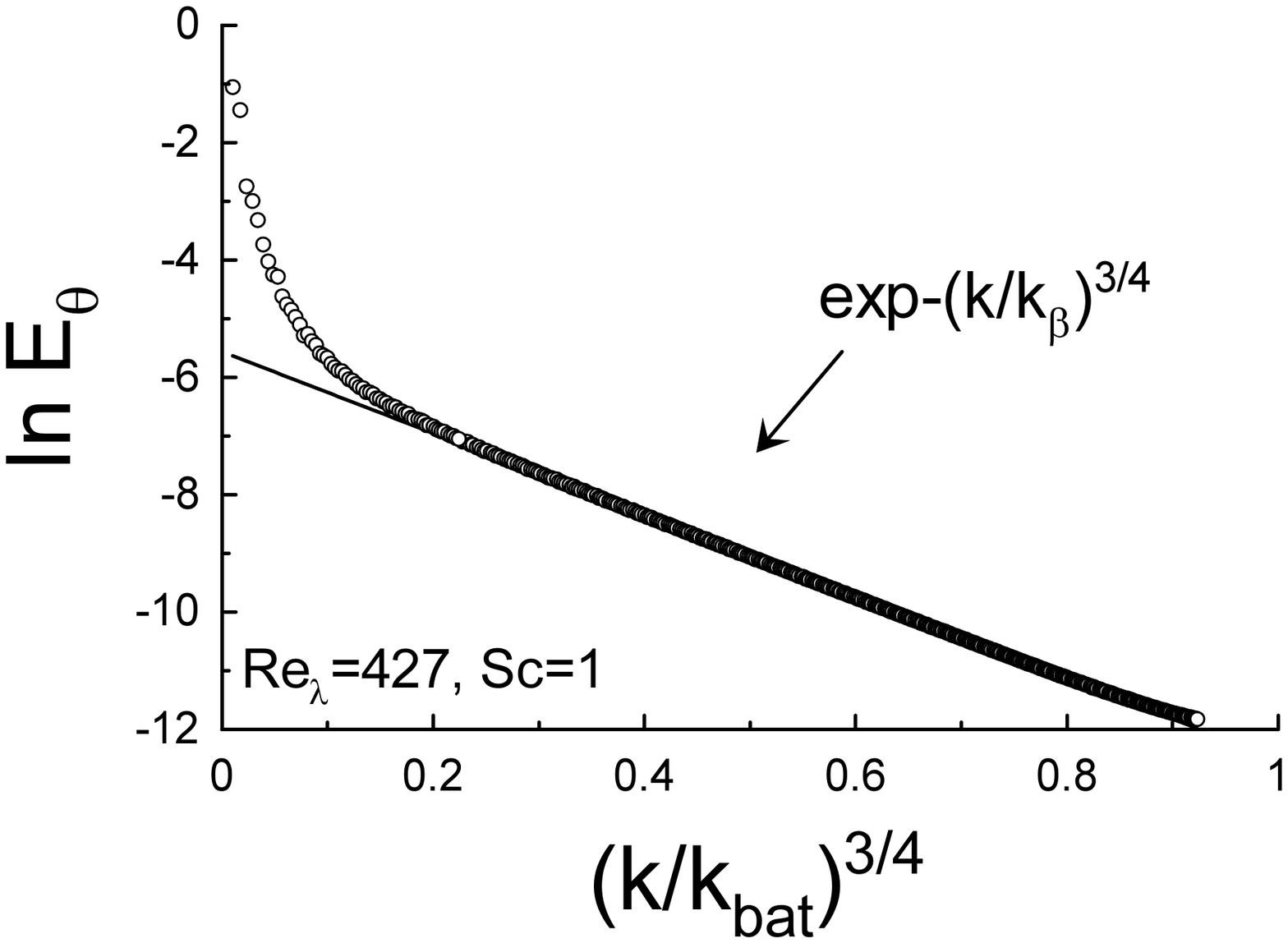}\vspace{-4cm}
\caption{\label{fig4} The same as in Fig. 1 but for three-dimensional power spectrum of a passive scalar from the DNS performed in Ref. \cite{wg} for $Re_{\lambda} = 427~~Sc=1$.}

\end{center}
\end{figure} 

\begin{figure}
\begin{center}
\includegraphics[width=8cm \vspace{-1.3cm}]{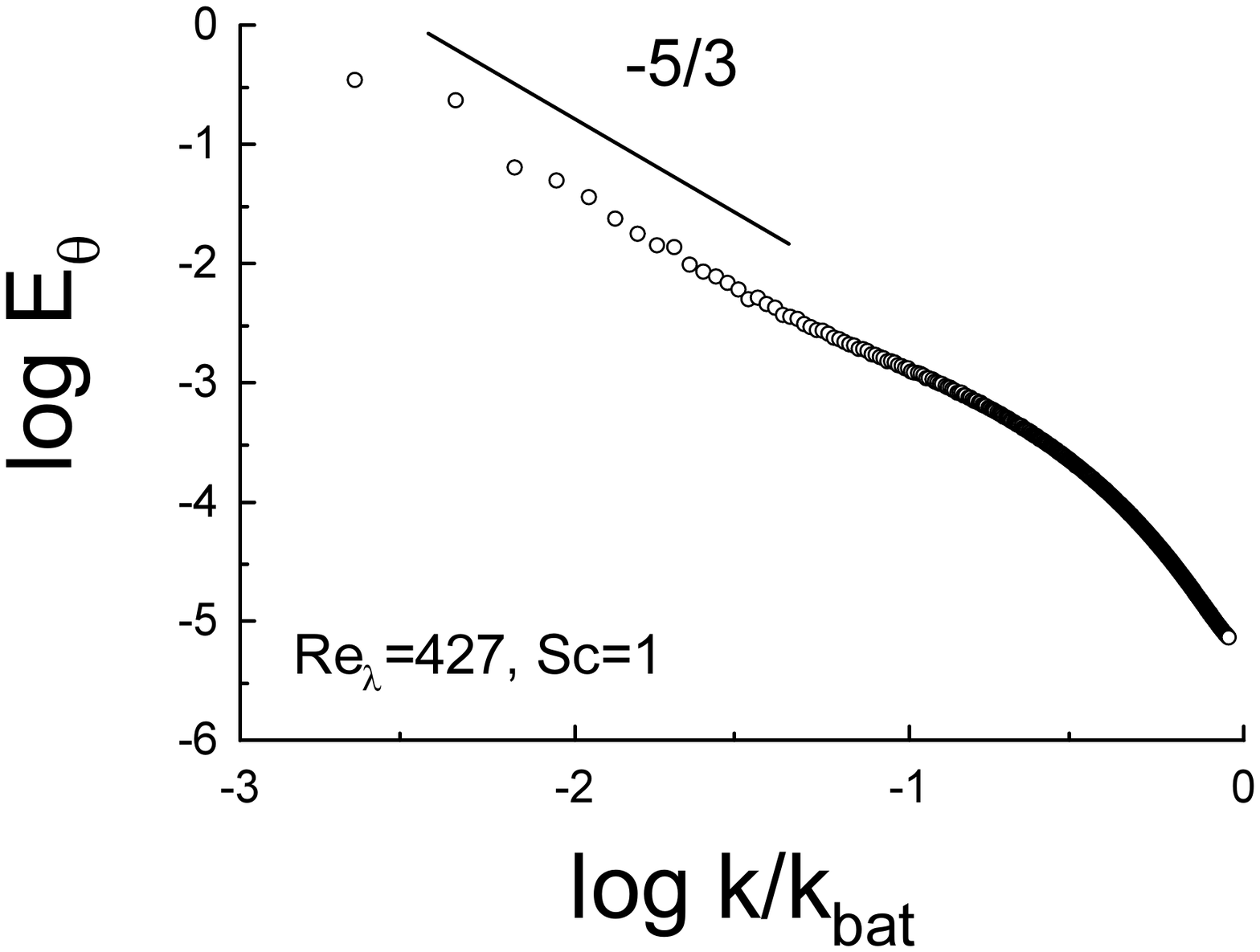}\vspace{-3.4cm}
\caption{\label{fig5} The same as in Fig. 4 but in the log-log scales}
\end{center}
\end{figure} 

\begin{figure}
\begin{center}
\includegraphics[width=8cm \vspace{-0.8cm}]{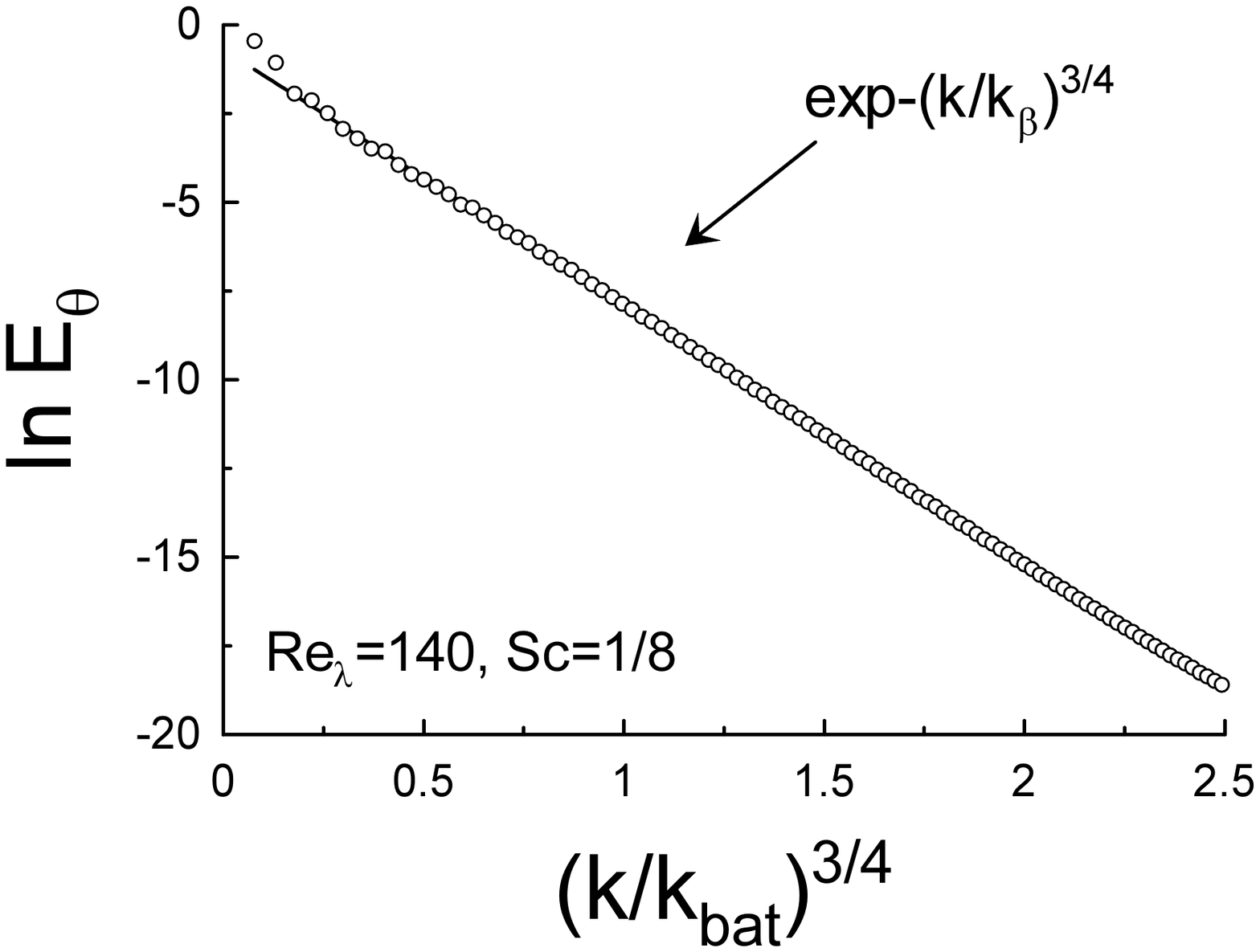}\vspace{-4cm}
\caption{\label{fig6} The same as in Fig. 1 but for three-dimensional power spectrum of a passive scalar from the DNS performed in Ref. \cite{yss} for $Re_{\lambda} = 140~~Sc=1/8$.} 
\end{center}
\end{figure} 

\begin{figure}
\begin{center}
\includegraphics[width=8cm \vspace{-1cm}]{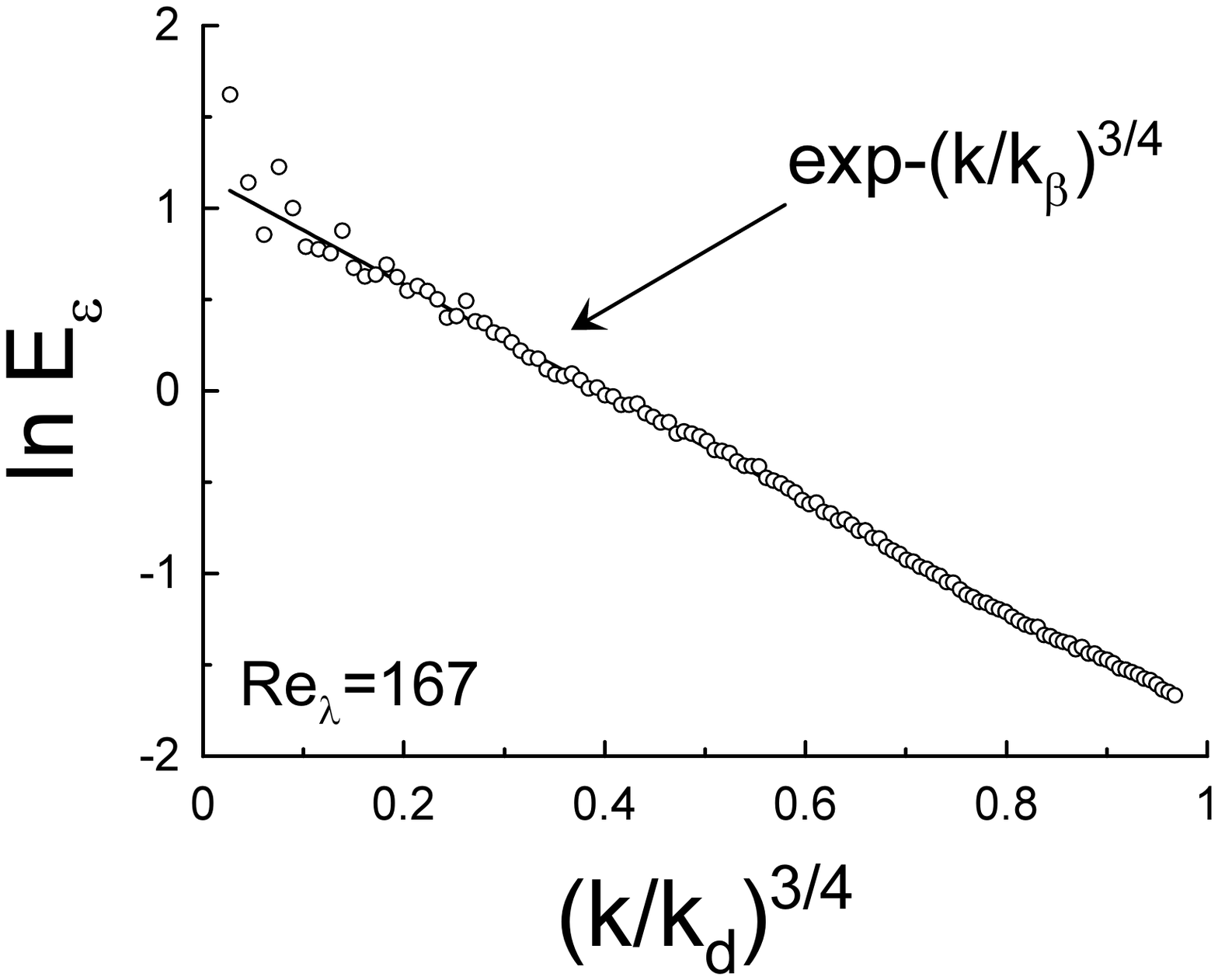}\vspace{-4cm}
\caption{\label{fig7} The same as in Fig. 1 but for normalized three-dimensional energy dissipation rate spectrum from the DNS performed in Ref. \cite{ishi} for Reynolds number $Re_{\lambda} = 167$.}
\end{center}
\end{figure}

\begin{figure}
\begin{center}
\includegraphics[width=8cm \vspace{-1cm}]{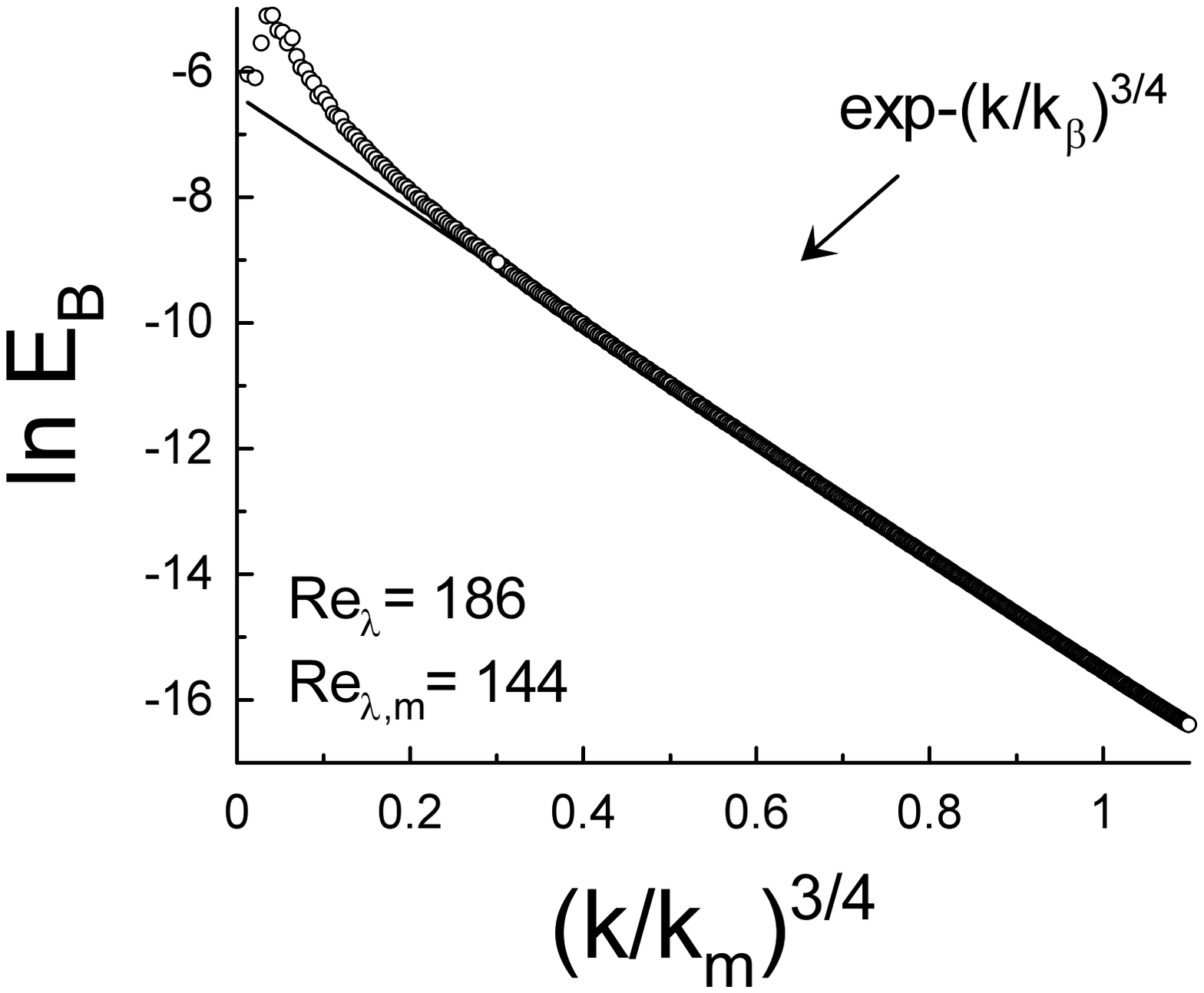}\vspace{-4cm}
\caption{\label{fig8} The same as in Fig. 1 but for three-dimensional power spectrum of the magnetic field. The data were taken from the DNS reported at the site http://turbulence.pha.jhu.edu/datasets.aspx (see also the Ref. \cite{a}).}
\end{center}
\end{figure}

Naturally, one can expect that the same spectrum should be also valid for a passive scalar field in the isotropic homogeneous turbulence. However, the Batchelor scale $k_{bat} = (\langle \varepsilon \rangle/\nu D^2)^{1/4}$ (where $D$ is the molecular diffusivity \cite{my}) is more relevant here than the Kolmogorov scale $k_d$. The scale $k_{\beta}$ in the terms of $k_{bat}$ will be independent not only on $Re_{\lambda}$ but also on the Schmidt number $Sc=\nu/D$ (for $Sc=1:~~$ $k_d=k_{bat}$ ). 

  Figure 4 shows the three-dimensional power spectrum of a passive scalar for $Re_{\lambda} =427$ and $Sc=1$. The data were taken from the Ref. \cite{wg} where the results of a DNS of a passive scalar mixing in the homogeneous isotropic steady three-dimensional turbulence are reported. The value of $k_{\beta}$ extracted from the straight line in Fig. 4 has the value $~k_{\beta} \simeq 0.075 k_{bat}$. In order to show emergence of the inertial (Obukhov-Corrsin) range adjacent to the range of the distributed chaos with the increase of $Re_{\lambda}$ we show in figure 5 the same data as in Fig. 4 but in the log-log scales (where the power-law spectrum corresponds to a straight line, cf. with Figs. 2 and 3).
Figure 6 shows the three-dimensional power spectrum of a passive scalar for $Re_{\lambda} =140$ and $Sc=1/8$. The data were taken from the Ref. \cite{yss} where the results of a DNS of a passive scalar mixing in the homogeneous isotropic steady three-dimensional turbulence are reported. The straight line is drawn in this figure in order to indicate the Eq. (7) with $\beta =3/4$. The stretched exponential covers about entire available data range.  The scale $k_{\beta}$ represented in the terms of $k_{bat}$ is $~k_{\beta} \simeq 0.072 k_{bat}$ (cf. with $~k_{\beta} \simeq 0.075 k_{bat}$ for the case: $Re_{\lambda} =427$ and $Sc=1$). \\

 The power spectrum of the energy dissipation rate 
 $\varepsilon ({\bf x})= \nu \frac{\partial u_i}{\partial x_k}\frac{\partial u_i}{\partial x_k} $ 
 can be also considered as a subject for this universal approach. Figure 7 shows normalized (non-dimensional) three-dimensional energy dissipation rate power spectrum $E_{\varepsilon}$ from a high-resolution direct numerical simulation of
homogeneous steady three-dimensional isotropic turbulence \cite{ishi} for
Reynolds number $R_{\lambda}=167$. Value of $k_{\beta} \simeq 0.234k_d$ for the energy dissipation rate field.  \\

Finally let us consider power spectrum of the magnetic field fluctuations in a forced incompressible  MHD turbulence. In this DNS \cite{a} energy is injected by a Taylor-Green flow stirring force. The magnetic Prandtl number is unity. The Kolmogorov scale based on the averaged magnetic field dissipation rate $\langle \varepsilon_m \rangle$: $~~k_m = (\langle \varepsilon_m \rangle /\nu^3)^{1/4}$. Taylor-scale Reynolds number for the velocity filed $~R_{\lambda} = 186$ , and for the magnetic field $~R_{\lambda , m}= 144$. The data are available at the site http://turbulence.pha.jhu.edu/datasets.aspx (the data provenance: H. Aluie, G. Eyink, E. Vishniac, S. Chen). Figure 8 shows the magnetic field power spectrum $E_{B}$ obtained in this DNS.  The scales in this figure are chosen in order to represent the Eq. (7) with $\beta =3/4$ as a straight line. The straight line is drawn in this figure in order to indicate the Eq. (7) with $\beta =3/4$. The value of $k_{\beta}$ extracted from the straight line in Fig. 8 has the same value $~k_{\beta} \simeq 0.052 k_m$ (in the terms of $k_m$) as for the velocity field  (in the terms of $k_d$) in the above considered isotropic turbulence (see also Figs. 1 and 2 and the above discussed theoretical estimate $~\ln(k_m/k_{\beta}) \simeq 3$).

\section{Acknowledgement}

I thank H. Aluie, S. Chen, G. Eyink, T. Ishihara, T. Nakano, D. Fukayama, T. Gotoh, K. R. Sreenivasan and E. Vishniac for sharing 
their data.

\end{document}